\documentclass[a4paper,11pt]{article}
\pdfoutput=1 % if your are submitting a pdflatex (i.e. if you have
\usepackage{jinstpub} % for details on the use of the package, please
\usepackage{hyperref}
\usepackage{subcaption}

\title{\boldmath Highly Granular Calorimeters: Technologies and Results}

\author[a,b,1]{Yong Liu,\note{on behalf of the CALICE collaboration}}

\affiliation[a]{Insitut f{\"u}r Physik, Johannes Gutenberg-Universit{\"a}t Mainz, Staudinger Weg 7, 55128 Mainz, Germany}
\affiliation[b]{Cluster of Excellence PRISMA Detector Lab, Johannes Gutenberg-Universit{\"a}t Mainz, Staudinger Weg 9, 55128 Mainz, Germany}

\emailAdd{yong.liu@uni-mainz.de}

\abstract{The CALICE collaboration is developing highly granular calorimeters for experiments at a future lepton collider primarily to establish technologies for particle flow event reconstruction. These technologies also find applications elsewhere, such as detector upgrades for the LHC. Meanwhile, the large data sets collected in an extensive series of beam tests have enabled detailed studies of the properties of hadronic showers in calorimeter systems, resulting in improved simulation models and development of sophisticated reconstruction techniques. In this proceeding, highlights are included from studies of the structure of hadronic showers and results on reconstruction techniques for imaging calorimetry. In addition, current R\&D activities within CALICE are summarized, focusing on technological prototypes that address challenges from full detector system integration and production techniques amenable to mass production for electromagnetic and hadronic calorimeters based on silicon, scintillator, and gas techniques.}


\proceeding{Instrumentation for Colliding Beam Physics 2017 (INSTR17)\\
  27 February - 3 March, 2017\\
  Novosibirsk, Russia}

\begin{document}
\maketitle
\flushbottom

\section{Introduction}
\label{sec:intro}

Driven by the precision physics programmes at future lepton collider experiments, e.g. the proposed International Linear Collider (ILC), it is essential to separate hadronic decays of W and Z bosons, which requires to achieve an unprecedented jet energy resolution of 3-4\% in a wide range of jet energies (e.g. 40-500 GeV at the ILC). The most promising strategy to achieve this goal is the particle flow approach~\cite{PFA}, which imposes requirements for highly granular calorimetry and also a compact calorimetry layout within the magnet coil. The CALICE collaboration are developing various options for the highly granular electromagnetic and hadronic calorimeters. At the first stage, physics prototypes were developed and exposed to test beams to validate the physics performance. Large data sets collected at test beams also provide possibilities to study shower properties and to test simulation models with high precision thanks to the calorimeters finely segmented in both transverse and longitudinal dimensions. As the next step, technological prototypes are being developed to address the challenges of scalability to build a final detector with around ten million readout units. Automated mass assembly will be indispensable to reach this goal.

In this proceeding, CALICE physics prototypes and highlights of shower studies will be shown in the Section~\ref{sec:physprot}. Progress on the development of CALICE technological prototypes will be covered in the Section~\ref{sec:techprot}, followed by other applications of CALICE-developed technologies in the Section~\ref{sec:otherapp}. A summary will be finally given in the Section~\ref{sec:sum}.

\section{CALICE physics prototypes and selected test-beam results}
\label{sec:physprot}

%\begin{figure}[htbp]
%\centering 
%\includegraphics[width=.8\textwidth]{PFA_Calorimetry_Tech_Tree}
%\caption{\label{fig:PfaCaloTree} Technology tree of calorimeters based on the particle-flow approach}
%\end{figure}

%Sampling calorimetry technologies for CALICE physics prototypes are summarized in Figure~\ref{fig:PfaCaloTree}, among which different active material options have been investigated.
\subsection{Technology options}
Various technologies for CALICE physics prototypes have been investigated, among which different active material options are summarized as following.
 
\begin{itemize}
\item Silicon: Silicon sensors have promising advantages of building sensitive layers with a minimal thickness (typically less than 1mm) and of fine pixelisation for the desired high granularity.
\item Scintillator: Scintillator is a robust and mature option for calorimetry with the flexibility to choose the cell size and thickness. Scintillation light is collected by a photosensor, e.g. a silicon photomultiplier (SiPM), which is compact and resistant to magnetic fields.
\item Gas: Gas as a cost-effective medium to cover a large area can also be easily segmented to reach a very high granularity. Resistive Plate Chambers (RPCs) are a mature option. The drawback of energy fluctuations in gas can be eliminated by an approach of counting avalanche hits in segmented cells. With the 1-bit readout (digital), the energy can be accurately reconstructed as long the average number of hits per cell is close to one. To further improve the linearity and energy resolution in a higher energy region, the concept of 2-bit readout (semi-digital) is developed with thresholds applied for hit counting.
\end{itemize}

CALICE physics prototypes of electromagnetic calorimeters (ECAL) as well as hadronic calorimeters (HCAL) are briefly summarized as below.
\begin{itemize}
\item Silicon-ECAL~\cite{SiW_ECal}: 30 layers, $18\times18$ cm$^{2}$ active area per layer, $1\times1$ cm$^{2}$ cells (silicon pads)
\item Scintillator-ECAL~\cite{ScW_ECal}: 26 layers, $9\times9$ cm$^{2}$ active area per layer, $1\times4.5$ cm$^{2}$ cells (scintillator strips with SiPM readout)
\item Analogue HCAL~\cite{AHCAL_PhysProt}: 38 layers, $90\times90$ cm$^{2}$ active area per layer, minimal cell size $3\times3$ cm$^{2}$ (scintillator tiles with SiPM readout)
\item Digital HCAL~\cite{DHCAL}: 54 layers,  $1\times1$ m$^{2}$ active area per layer, $1\times1$ cm$^{2}$  cells (RPC pads with binary readout, i.e. each pad with a single threshold)
\item Semi-Digital HCAL~\cite{SDHCAL}: 50 layers, $1\times1$ m$^{2}$ active area per layer, $1\times1$ cm$^{2}$ cells (Glass RPC pads with 2-bit readout, i.e. each pad with 3 thresholds)
\end{itemize}

\subsection{Selected highlights of the performance of combined calorimeters}

\begin{figure}[htbp]
\centering 
\includegraphics[width=.8\textwidth]{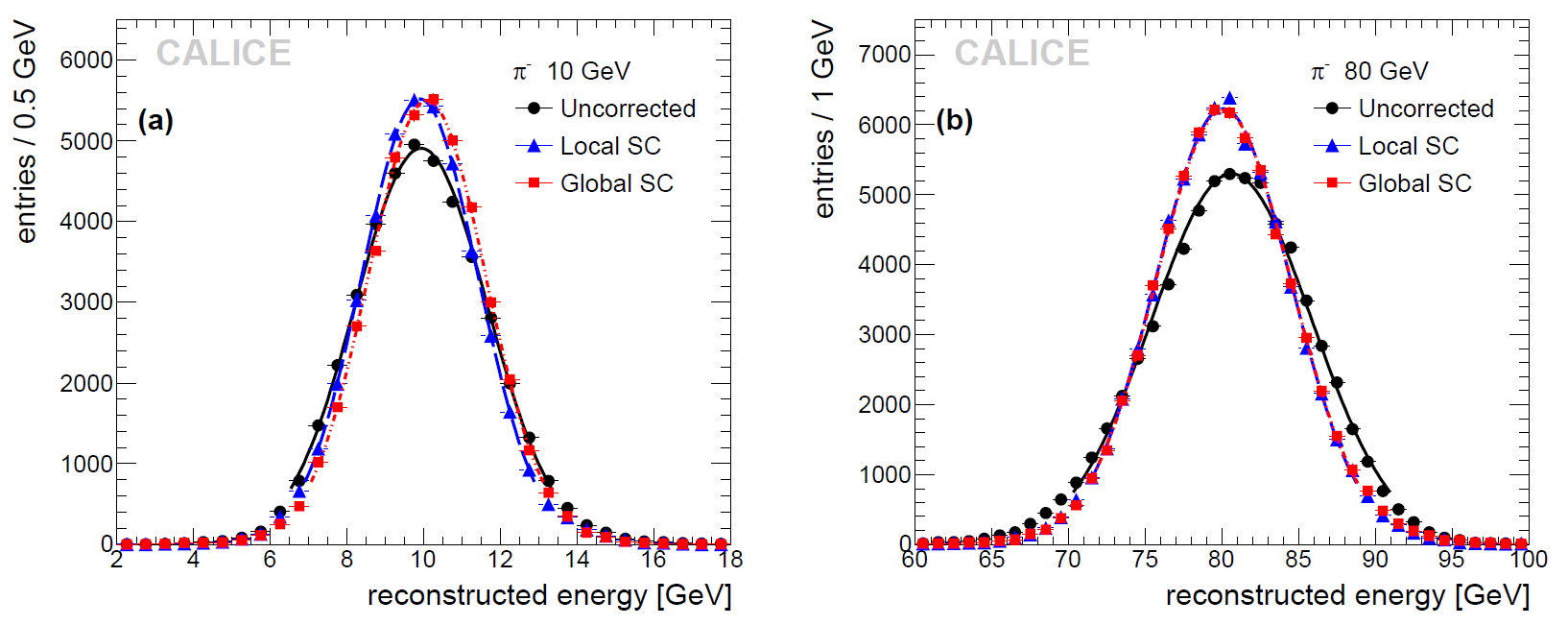}
\caption{\label{fig:PionEneRec} Reconstructed energy distributions of 10 GeV $\pi^{-}$ (a) and 80 GeV $\pi^{-}$ (b) without compensation (in black circles) and after local software compensation (SC) in blue triangles, and after global software compensation in red squares~\cite{AHCAL_SC}.}
\end{figure}

\begin{figure}[htbp]
\centering
\begin{minipage}{.48\textwidth}
  \centering
  \includegraphics[width=.9\linewidth]{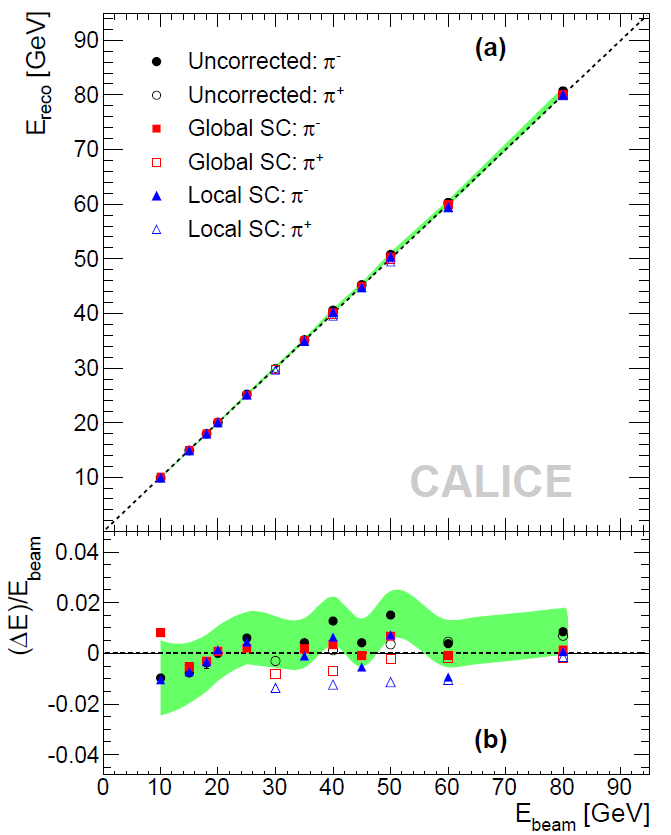}
  \caption{(a) Mean reconstructed energy for pions and (b) relative residuals to beam energy versus beam energy. Dotted lines correspond to $E_{\mathrm{reco}} = E_{\mathrm{beam}}$. Systematic uncertainties are indicated by the green band, corresponding to the uncertainties for the uncorrected $\pi^{-}$ data sample~\cite{AHCAL_SC}.}
  \label{fig:PionLinearity}
\end{minipage}%
\quad
\begin{minipage}{.48\textwidth}
\centering
\includegraphics[width=.8\linewidth]{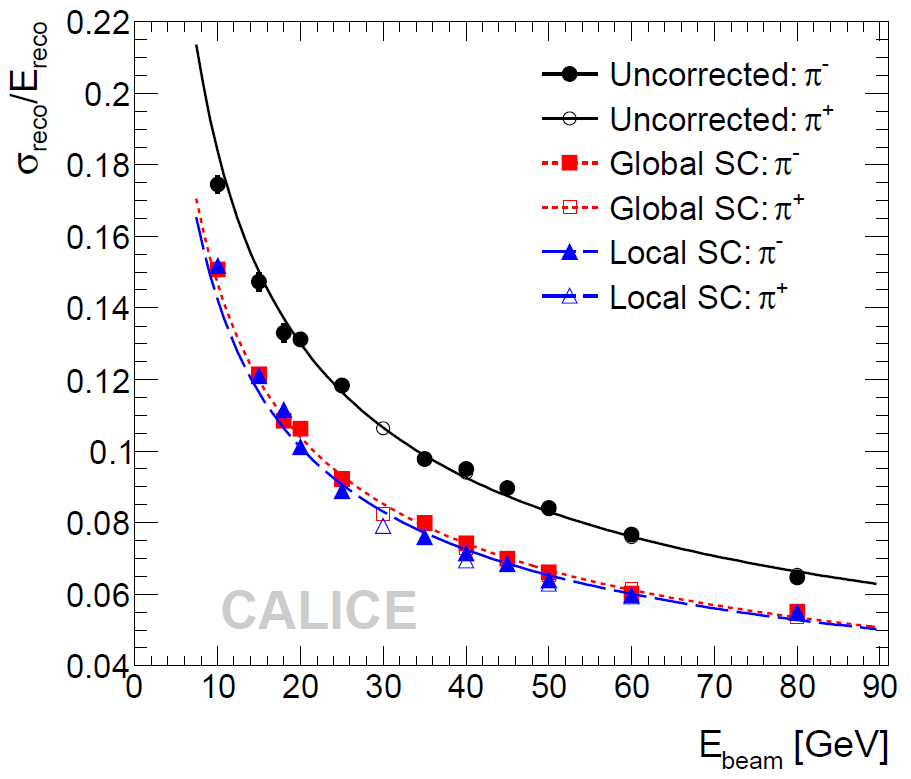}
\caption{Energy resolution versus beam energy~\cite{AHCAL_SC}.}
\label{fig:PionEneReso}
\centering
\includegraphics[width=.75\linewidth]{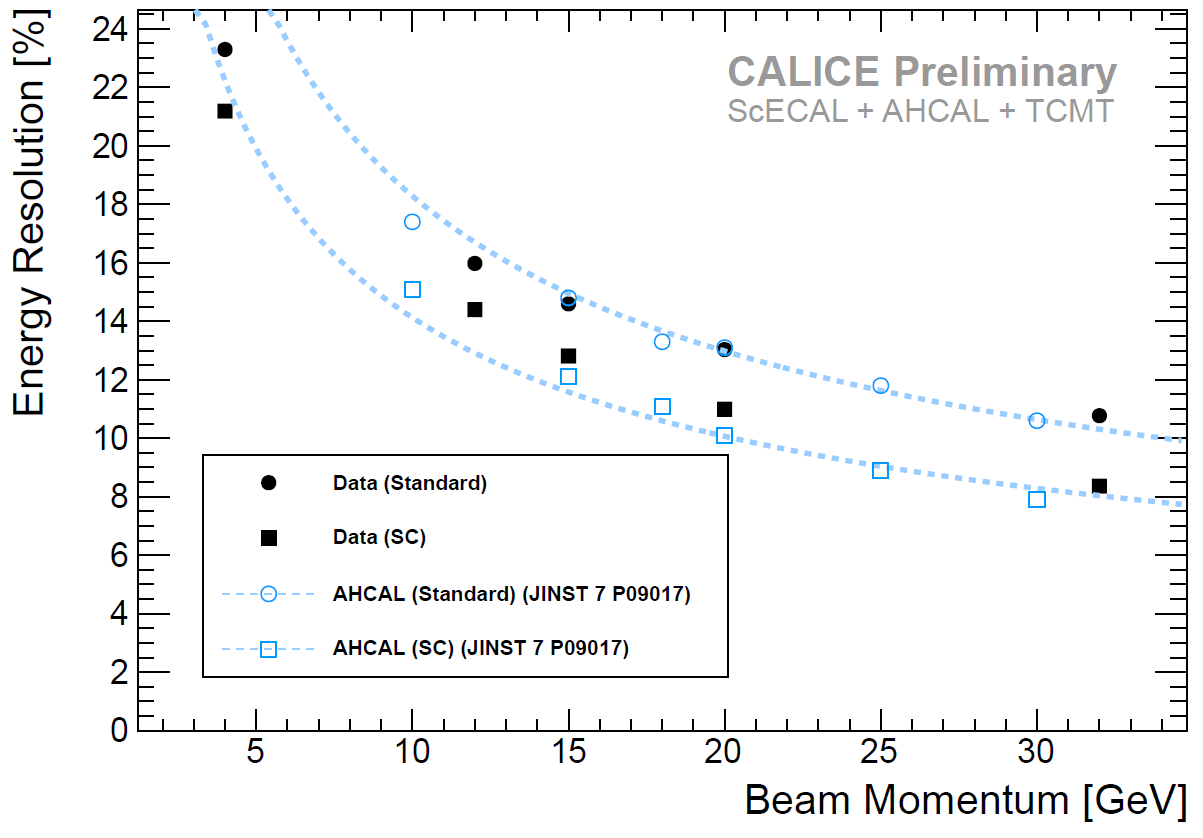}
\caption{Comparison of energy resolutions of two calorimeter systems using pion samples, with the only difference of two ECAL technologies~\cite{ScECAL+AHCAL}.}
\label{fig:PionEneResoComp}

\end{minipage}
\end{figure}

The hadronic energy resolution of the AHCAL physics prototype with steel absorbers has been studied using the pion data collected at the CERN SPS in 2007 in the energy range from 10 to 80 GeV and two software compensation techniques have been implemented using the information of the local energy density within showers obtained from the highly granular readout~\cite{AHCAL_SC}. Reconstructed energy distributions are shown in Figure~\ref{fig:PionEneRec}. Figure~\ref{fig:PionLinearity} shows the linearity of the calorimeter response to pions showering predominantly in the AHCAL, which is within $\pm1.5\%$ in the studied energy range. The fractional energy resolution ($\sigma_{\mathrm{reco}}/E_{\mathrm{reco}}$) is shown in Figure~\ref{fig:PionEneReso}. The hadronic energy resolution is improved by around 20~\% benefited from the software compensation, where the stochastic term is reduced from $58\%/\sqrt{E/\mathrm{GeV}}$ to $45\%/\sqrt{E/\mathrm{GeV}}$. %The stochastic term is $(57.6\pm0.4)\%$, $(45.8\pm0.3)\%$ and $(44.3\pm0.3)\%$, with constant terms of $(1.6\pm0.3)\%$, $(1.6\pm0.2)\%$ and $(1.8\pm0.3)\%$ for the uncorrected resolution, global software compensation and local software compensation, respectively.
Latest progress of pion data analysis, shown in Figure~\ref{fig:PionEneResoComp}, indicates that the performance of the combined energy reconstruction with two calorimeter systems only different in two ECAL technologies (i.e. Silicon-ECAL and Scintillator-ECAL, with different longitudinal sampling fractions) finally converge~\cite{ScECAL+AHCAL}.

\subsection{In-depth understanding of hadronic showers}
Precision measurements of hadronic showers using highly granular calorimeters also make it feasible to understand better hadronic showers and to finely test shower models.

\begin{figure}[htbp]
\centering
\begin{minipage}{.35\textwidth}
  \centering
  \includegraphics[width=.98\linewidth]{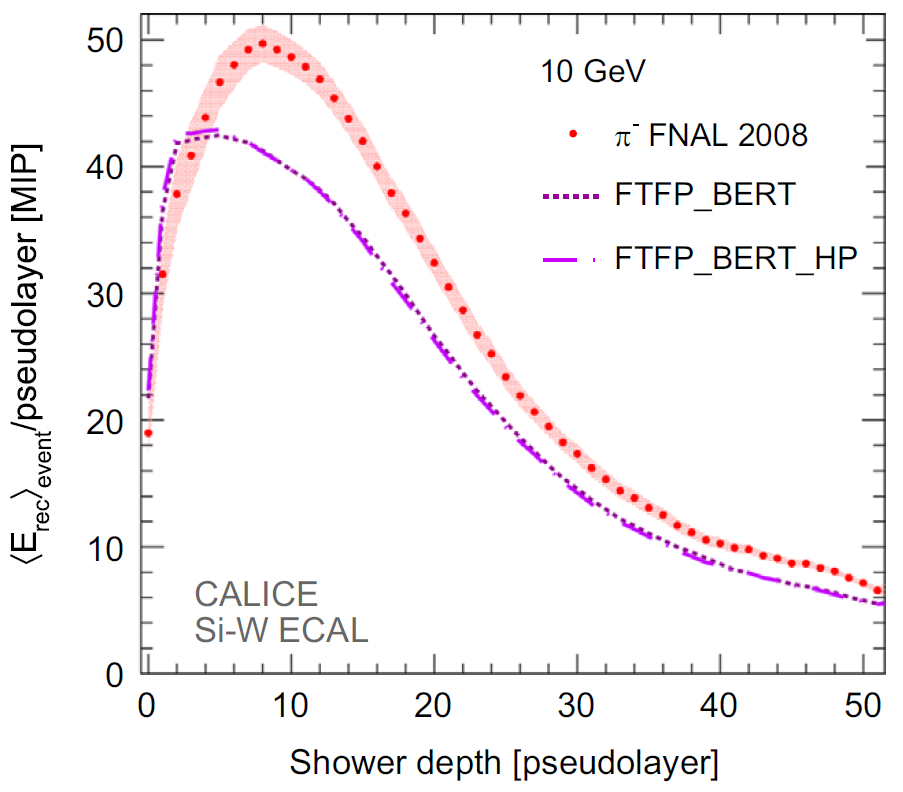}
  \caption{The longitudinal energy profile for interacting events at 10 GeV in the Silicon-Tungsten ECAL for data and simulation models~\cite{SiECAL_Pion}.}
  \label{fig:SiECAL_Pion}
\end{minipage}%
\quad
\begin{minipage}{.62\textwidth}
\centering
\includegraphics[width=.98\linewidth]{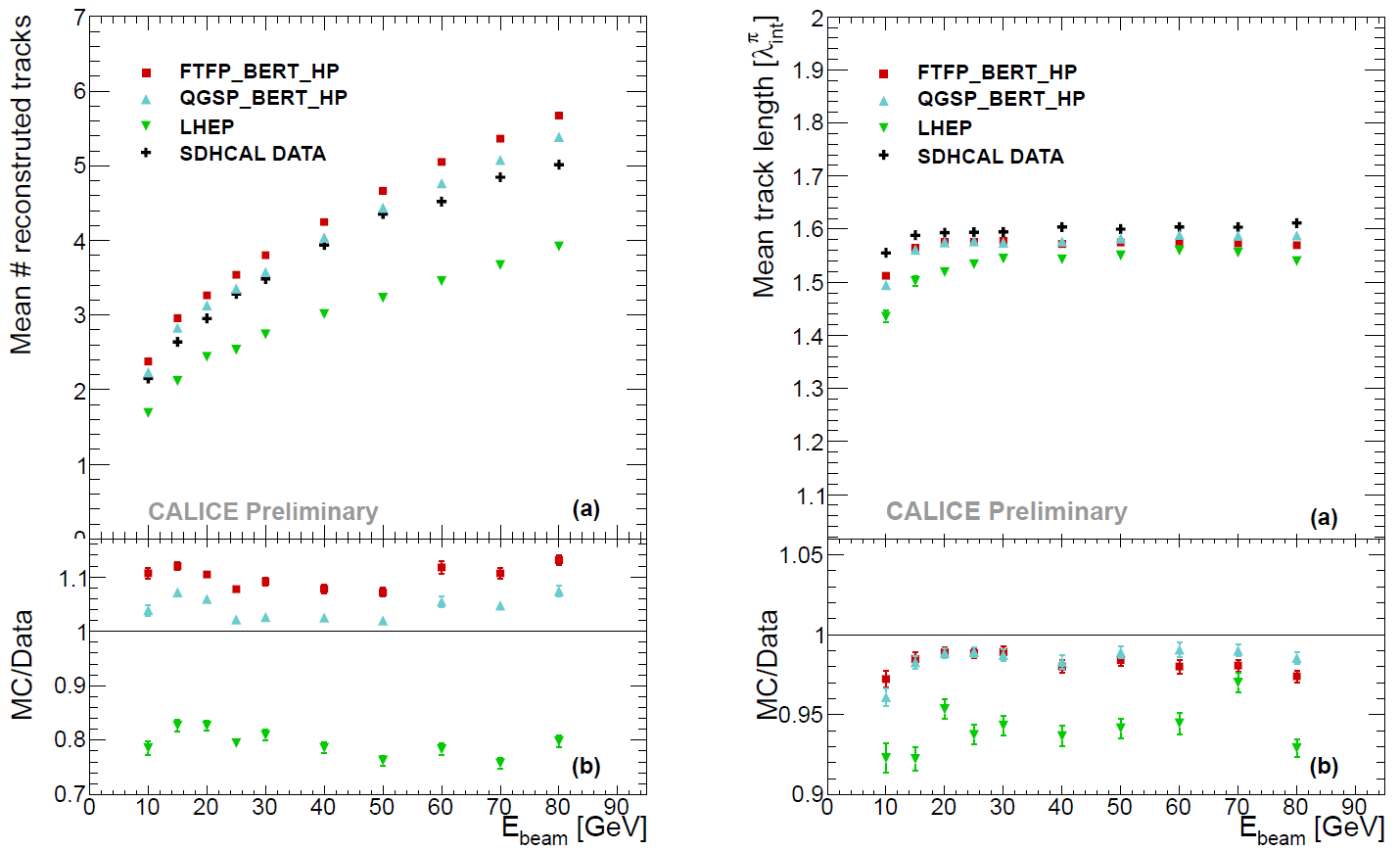}
\caption{Mean number of reconstructed tracks (left) and mean track length (right) in hadronic showers in the SDHCAL as a function of the beam energy~\cite{SDHCAL_MipTrack}.}
\label{fig:SDHCAL_MipTrack}
\end{minipage}
\end{figure}

The response of the Silicon-Tungsten ECAL to first hadronic interactions has been studied based on the data sample of negatively charged pions in the energy range of 2-10 GeV collected at FNAL in 2008~\cite{SiECAL_Pion}. Figure~\ref{fig:SiECAL_Pion} shows the shower profile of pions at 10 GeV, where the existing simulation models significantly deviate from the measured deposited energy.

Hadronic showers are being studied with the SDHCAL based on the pion data recorded at CERN SPS in 2012~\cite{SDHCAL_MipTrack}. The Hough transformation method is implemented to identify tracks of charged particles in the sparse parts (halos) of hadronic showers. This method proves to be highly efficient after the hits from dense cores in hadronic showers are eliminated. Figure~\ref{fig:SDHCAL_MipTrack} shows distributions of the mean number of reconstructed tracks and the mean track length at different energy points. Discrepancies between the data and hadronic models have been observed.

\begin{figure}[htbp]
\centering
\begin{minipage}{.48\textwidth}
  \centering
  \includegraphics[width=.98\linewidth]{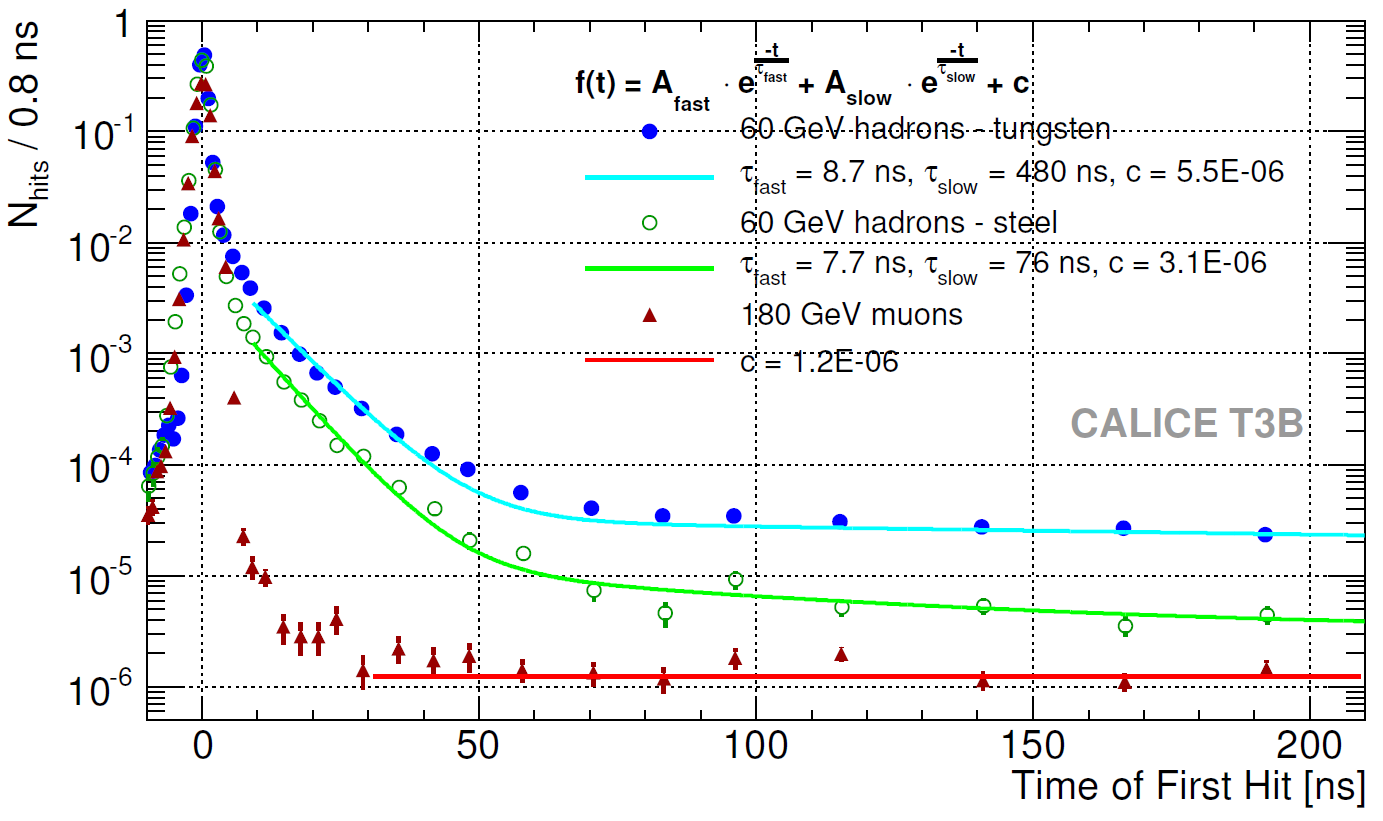}
  \caption{Time of first hit distribution of muon data with steel absorbers and hadron data with steel and tungsten absorbers in the T3B detector~\cite{T3B_Detector} with steel and tungsten absorbers in a time range of -10ns to 200ns~\cite{T3B}}.
  \label{fig:T3B_FirstHits}
\end{minipage}%
\quad
\begin{minipage}{.48\textwidth}
\centering
\includegraphics[width=.98\linewidth]{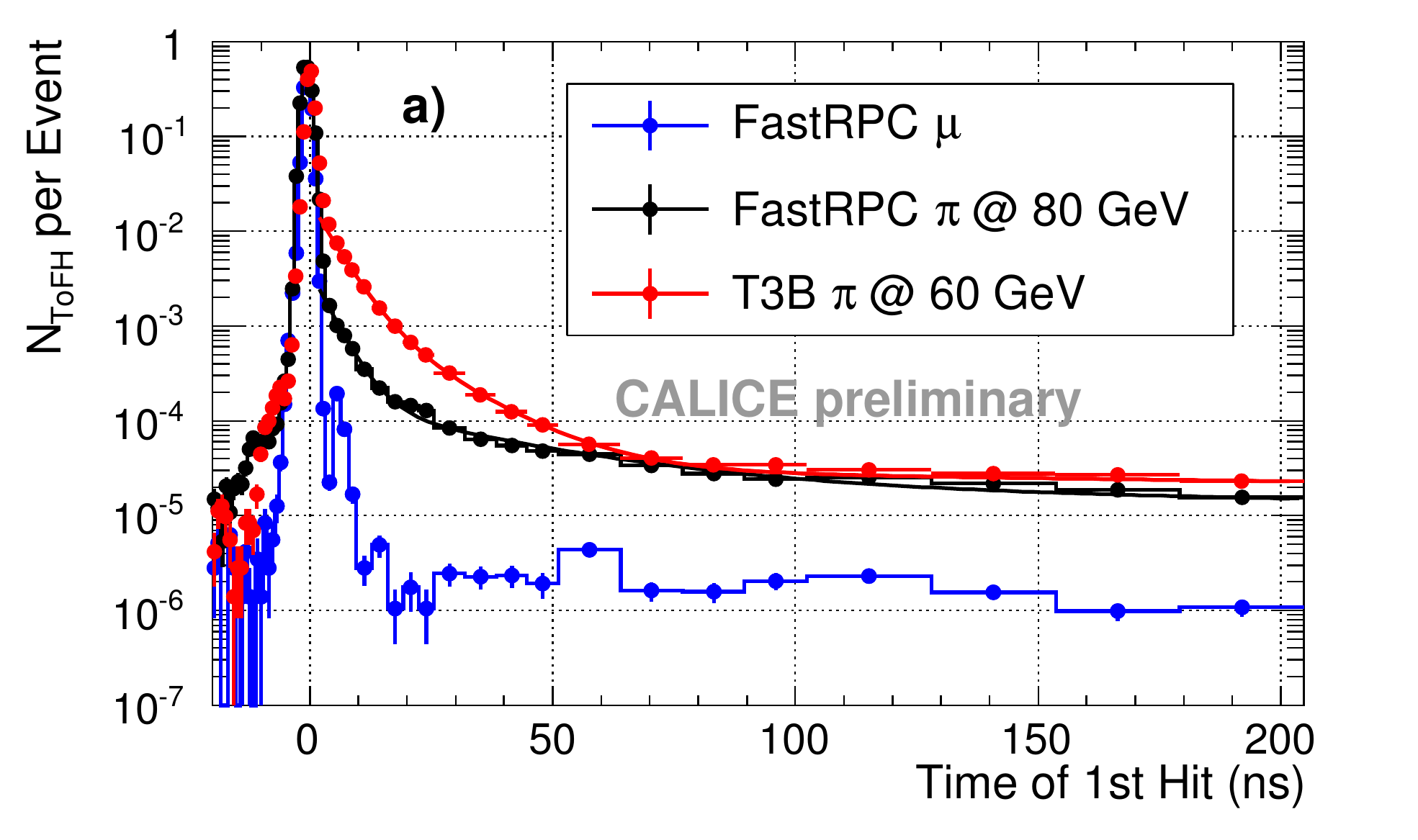}
\caption{Comparison of the time of first hit distribution observed in FastRPC (80 GeV $\pi^{\pm}$) and T3B (60 GeV $\pi^{\pm}$), normalized to the number of events with at least one hit in the FastRPC and T3B detectors, respectively~\cite{T3B_RPC}.}
\label{fig:ShowerTiming_ScVsGas}
\end{minipage}
\end{figure}

The time structure of hadronic showers in different active and absorber materials has been extensively studied with the T3B detector using scintillator tiles individually read out by SiPMs~\cite{T3B_Detector}. Figure~\ref{fig:T3B_FirstHits} shows the time profiles of different particles in the T3B detector. Muons give prompt responses; hadron showers exhibit an instantaneous component, followed by a fast shower component at intermediate times, and later a slower component starts smoothly to dominate. The late shower component shows more prominent in tungsten than in steel, which is consistent with the expectation that more neutrons are released in tungsten leading to delayed energy depositions. Figure~\ref{fig:ShowerTiming_ScVsGas} shows that scintillator tiles in the T3B detector are more sensitive than the gas-based RPC to the MeV-scale neutrons generated in the tungsten absorbers.

\section{CALICE technological prototypes}
\label{sec:techprot}
Recent progress on the development of technological prototypes based on gas and scintillator will be discussed; the Silicon-ECAL technological prototype will be introduced in a separate proceeding.

\subsection{Scintillator-tungsten ECAL technological prototype}
\begin{figure}[htbp]
\centering
\begin{minipage}{.35\textwidth}
  \centering
  \includegraphics[width=.98\linewidth]{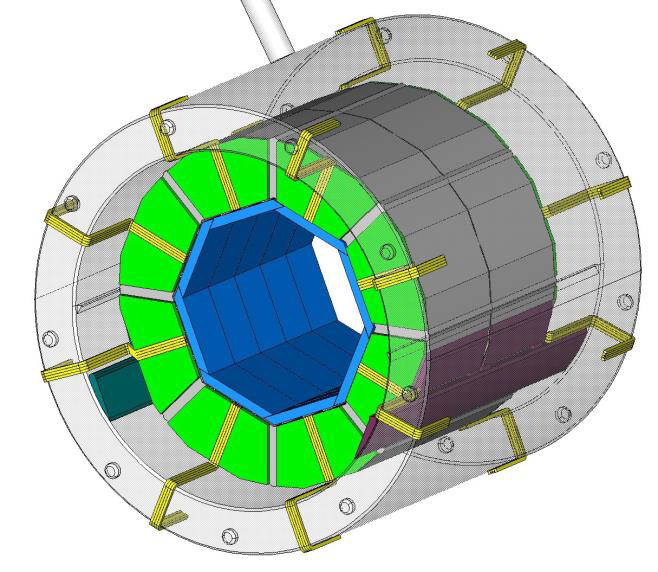}
  \caption{The schematics of an option for the ILC barrel calorimeter system (i.e. ECAL in blue and HCAL in green) within the magnet coil (in light grey).}
  \label{fig:ILD_Calo}
\end{minipage}%
\quad
\begin{minipage}{.62\textwidth}
\centering
\includegraphics[width=.98\linewidth]{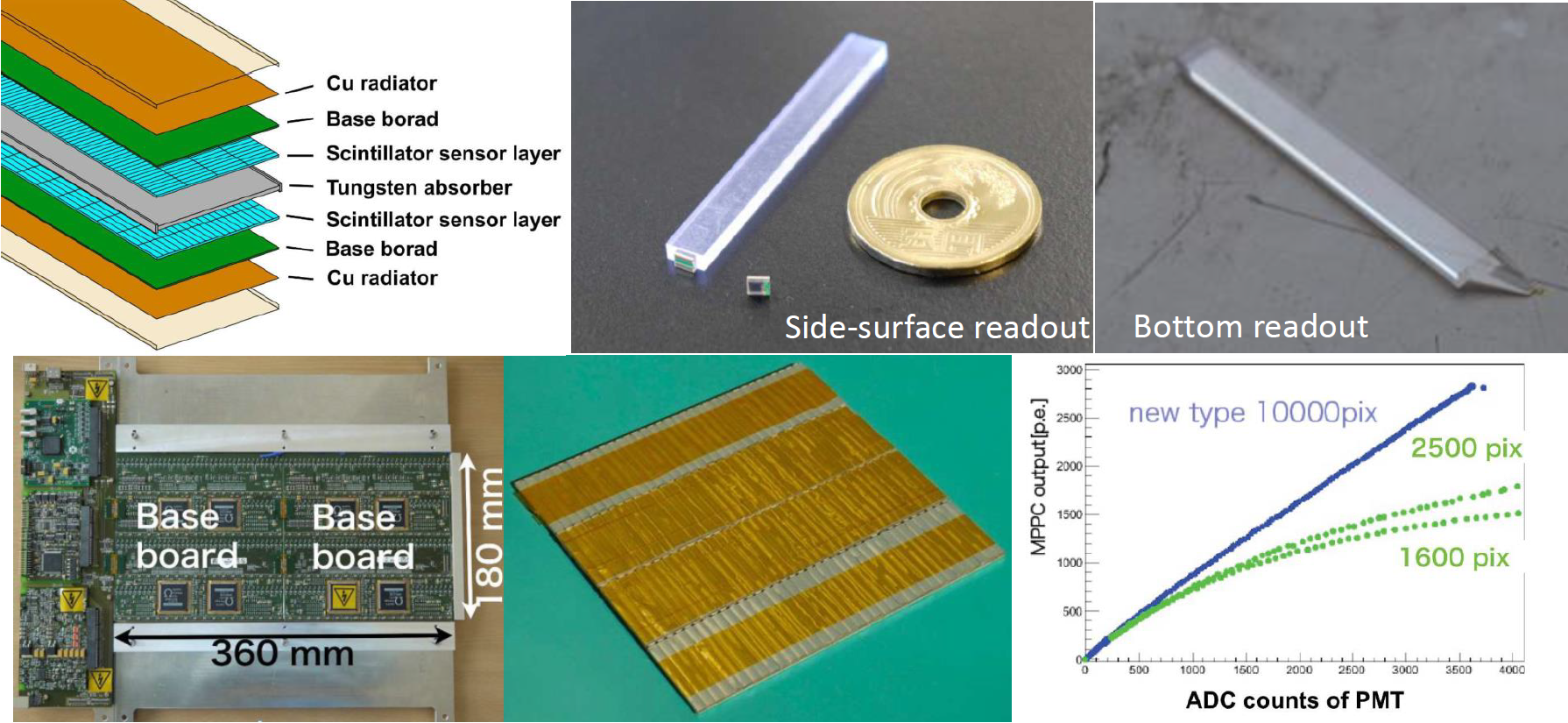}
\caption{Schematics of the structure of the Scintillator-ECAL (top left), 2 options of active components (top middle and right), a fully integrated readout board equipped with active components (bottom left and middle) and comparison of the response linearity of different SiPMs (bottom right). }
\label{fig:ScECAL_TechProt}
\end{minipage}
\end{figure}

A proposed layout of the compact barrel ECAL system at the ILC is shown as the inner wedges (in blue) in Figure~\ref{fig:ILD_Calo}. Active layers of scintillator-ECAL technological prototype have been improved in several aspects with highlights shown in Figure~\ref{fig:ScECAL_TechProt}. Wavelength-shifting fibres have been eliminated, thus SiPMs can directly coupled with scintillator strips ($45\times5\times2\mathrm{mm^3}$ or even thinner $45\times5\times1\mathrm{mm^3}$). SiPMs in surface-mounted packages (SMD-SiPMs) are being considered for easier assembly; strips are optimized by introducing tapered wedges for the bottom readout and for better response uniformity. SiPMs with very small pixels (e.g. $10\times10\mathrm{{\mu}m^2}$ per pixel, in total 10000 pixels in the $1\times1\mathrm{mm^2}$ sensitive area) are being investigated to reach a wider dynamic range. Strips and SiPMs are mounted directly onto the readout PCB, full integrated with front-end electronics. There have been smooth beam tests combined with the AHCAL technological prototype at CERN.

\subsection{AHCAL technological prototype}
\begin{figure}[htbp]
\centering
\begin{minipage}{.58\textwidth}
  \centering
  \includegraphics[width=.98\linewidth]{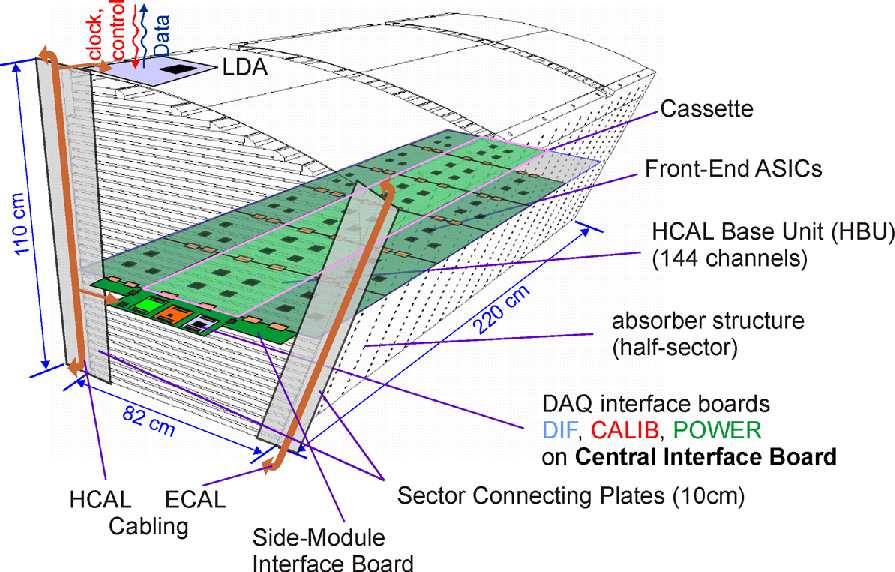}
  \caption{The schematics of a half octant of the final barrel AHCAL detector at the ILC. Active layers and readout electronics are fully integrated inside the absorber structure. DAQ interface boards as well as all the cabling for the combined calorimeter system are kept outside absorbers.} 
  \label{fig:AHCAL_TechProt}
\end{minipage}%
\quad
\begin{minipage}{.38\textwidth}
  \centering
  \includegraphics[width=.98\linewidth]{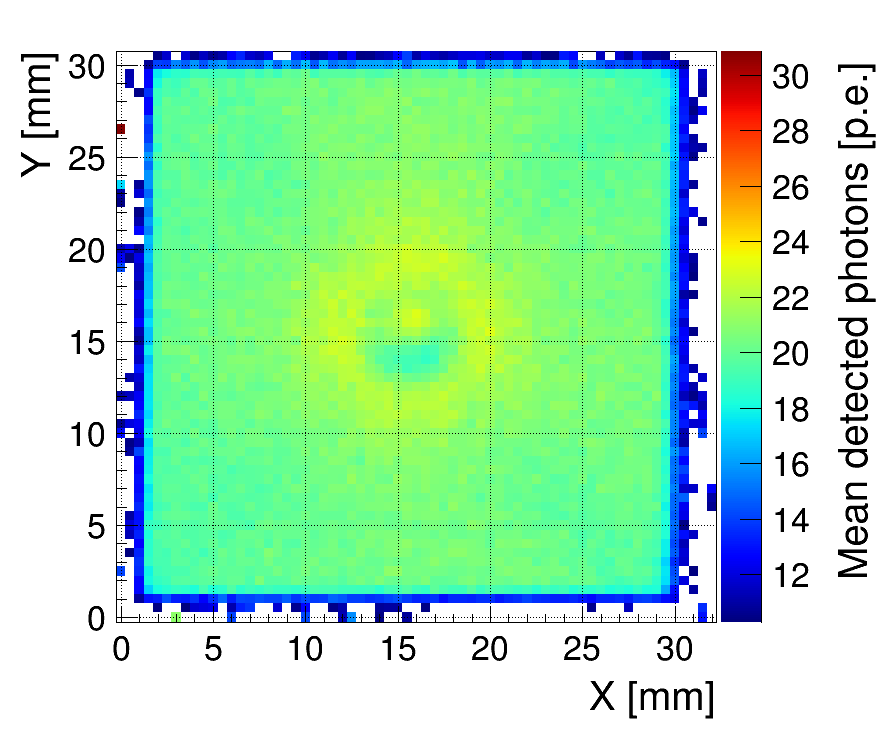}
  \caption{The uniformity map of a scintillator tile coupled with an SMD-SiPM in the centre of the tile bottom surface, obtained using a Sr-90 beta source~\cite{NovelTileDesign}. The x and y values of each point show where the incident electrons pass through the tile and the colour scale represents the mean response of the SMD-SiPM.} 
  \label{fig:TileUniformity}
\end{minipage}
\end{figure}

\begin{figure}[htbp]
\centering
\includegraphics[width=.9\linewidth]{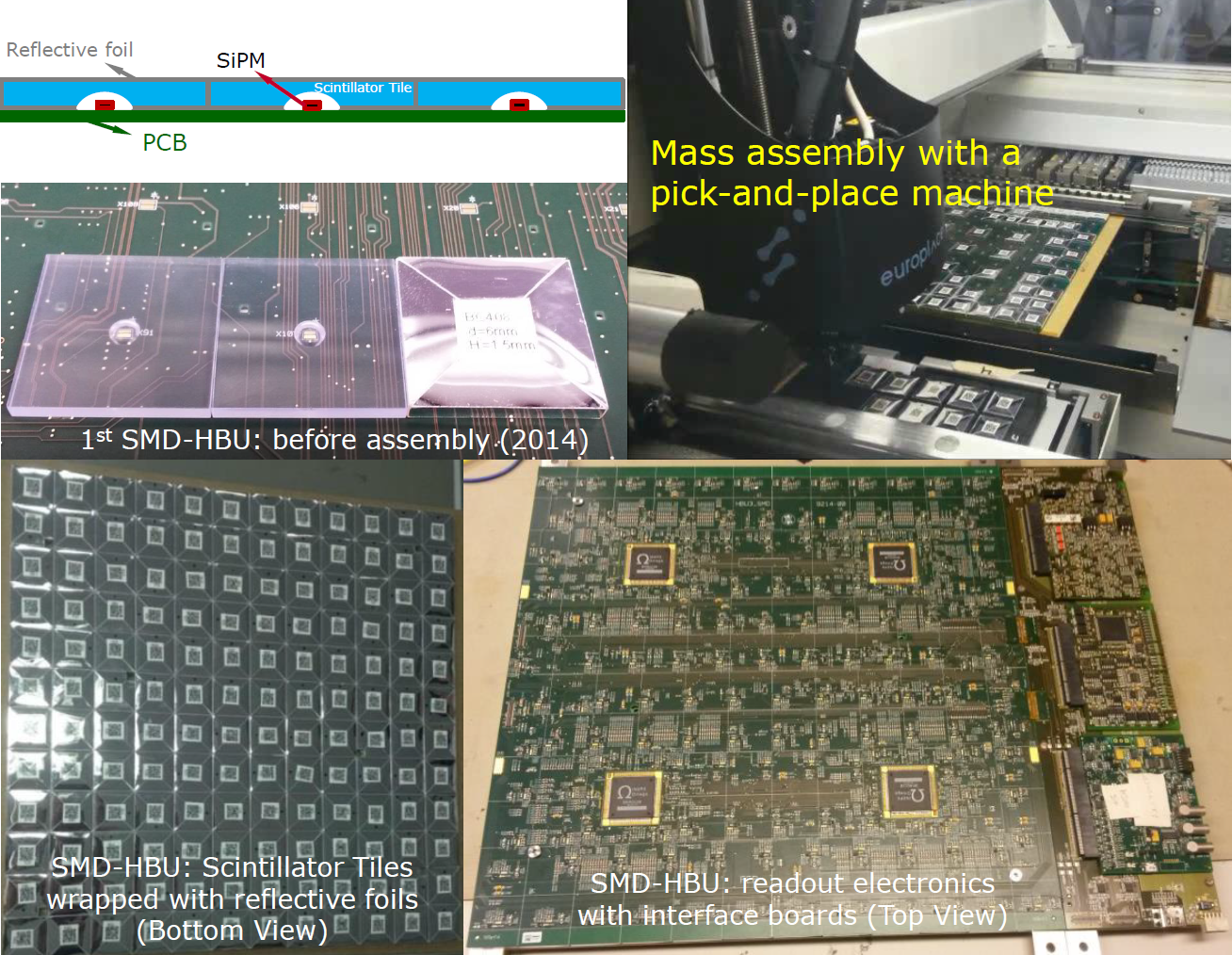}
\caption{Development of the first SMD-HBU from design to reality, including the schematics of the SMD-HBU and pioneering activities of the mass assembly of the first SMD-HBU using a pick-and-place matchine.}
\label{fig:SMD-HBU_Mainz_2014}
\end{figure}

The layout of the barrel AHCAL at the ILC is shown in the outer wedges (in green) in Figure~\ref{fig:ILD_Calo}. Figure~\ref{fig:AHCAL_TechProt} shows the schematics of the AHCAL technological prototype. A novel design of scintillator tiles individually read out by SMD-SiPMs has been developed, which makes it feasible to build the final AHCAL detector with in total $10^7$ channels via automated mass assembly~\cite{NovelTileDesign}. The design has been optimised by a Geant4-based full simulation model and has achieved promising performance in various tests. Figure~\ref{fig:TileUniformity} shows the excellent response uniformity obtained with electrons passing through the tile at different positions. The first AHCAL readout board (SMD-HBU) using this tile design with 144 channels was successfully built via mass assembly, as shown in Figure~\ref{fig:SMD-HBU_Mainz_2014}.

\begin{figure}[htbp]
\centering 
\includegraphics[width=.9\textwidth]{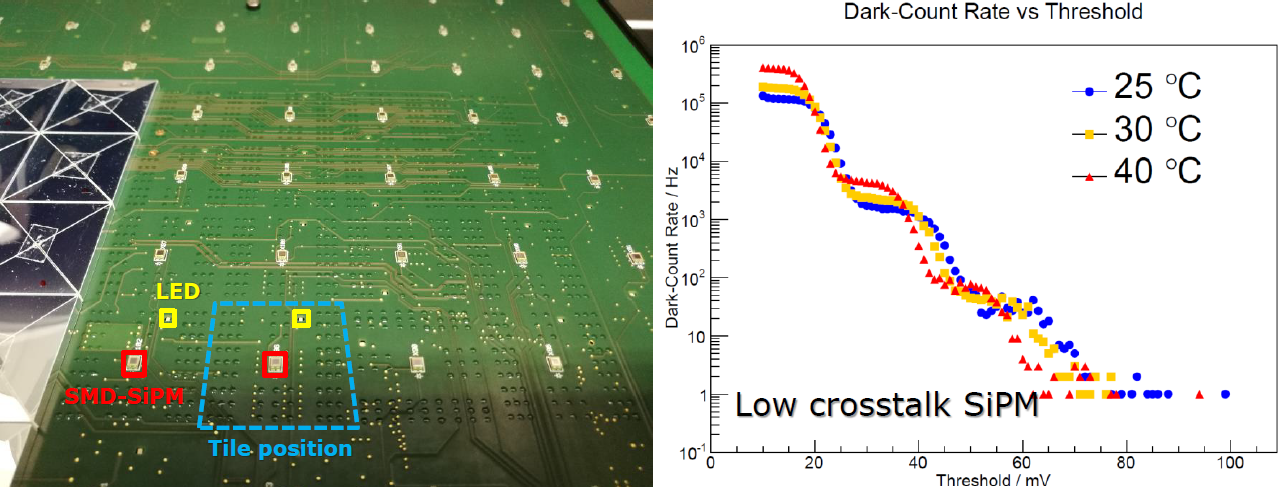}
\caption{\label{fig:SMD-HBU_Mainz_2016} Highlights of the assembly process of new AHCAL readout boards (SMD-HBUs with glue dispensed on the PCB (left); and results of the new SMD-SiPM of the dark-count rates at different thresholds at 3 typical temperatures (right).}
\end{figure}

The state-of-art SiPM now has very low inter-pixel crosstalk, which allows minimum noises in the AHCAL detector. A sample of this type has been characterised in a climate cabinet at different temperatures, as shown in Figure~\ref{fig:SMD-HBU_Mainz_2016} (top right). As a rehearsal step of building the AHCAL technological prototype, 6 new SMD-HBUs have been assembled using an improved procedure of mass assembly and an updated tile design, as shown in Figure~\ref{fig:SMD-HBU_Mainz_2016}. All 7 SMD-HBUs have shown promising performance in extensive tests using the cosmic-ray setup at Mainz as well as beam facilities at CERN and DESY. The novel tile design and the new type of SMD-SiPMs have been adopted as the baseline design for the AHCAL technological prototype. 

The AHCAL technological prototype will be instrumented with at least 40 active layers using a steel stack (an infrastructure for max.~48 layers), with each layer of $2\times2$ HBUs. Scintillator tiles have been produced via injection moulding and will be wrapped individually with reflective foils using a custom-made automated wrapping machine. Quality assurance setups have been developed and are being optimised, including a test setup of readout chips to test the functions of each chip, a delicate SiPM setup to check a few samples from each SiPM batch and a large cosmic-ray setup to measure simultaneously the performance of many HBUs in stack after the assembly.

In addition, progress has been made in the development of the DAQ hardware (Wing Link-Data Aggregator, or Wing-LDA) and software (EUDAQ). The Wing-LDA, developed for collecting data from all active layers of 2 neighbouring calorimeter segments, is compatible with the final compact calorimetry layout at the ILC. Electronics of the new SMD-HBU is being optimised for the power-pulsing mode at the ILC and extensive efforts have been invested in testing the power-pulsing performance of the AHCAL technological prototype using the electron beams at DESY. Further tests will also be carried out to check the performance in a strong magnetic field. After the prototype is fully assembled, it will also be commissioned using the hadron beams at CERN and also further precision studies on hadronic showers will be followed.

\subsection{SDHCAL technological prototype}

\begin{figure}[htbp]
\centering 
\includegraphics[width=.9\textwidth]{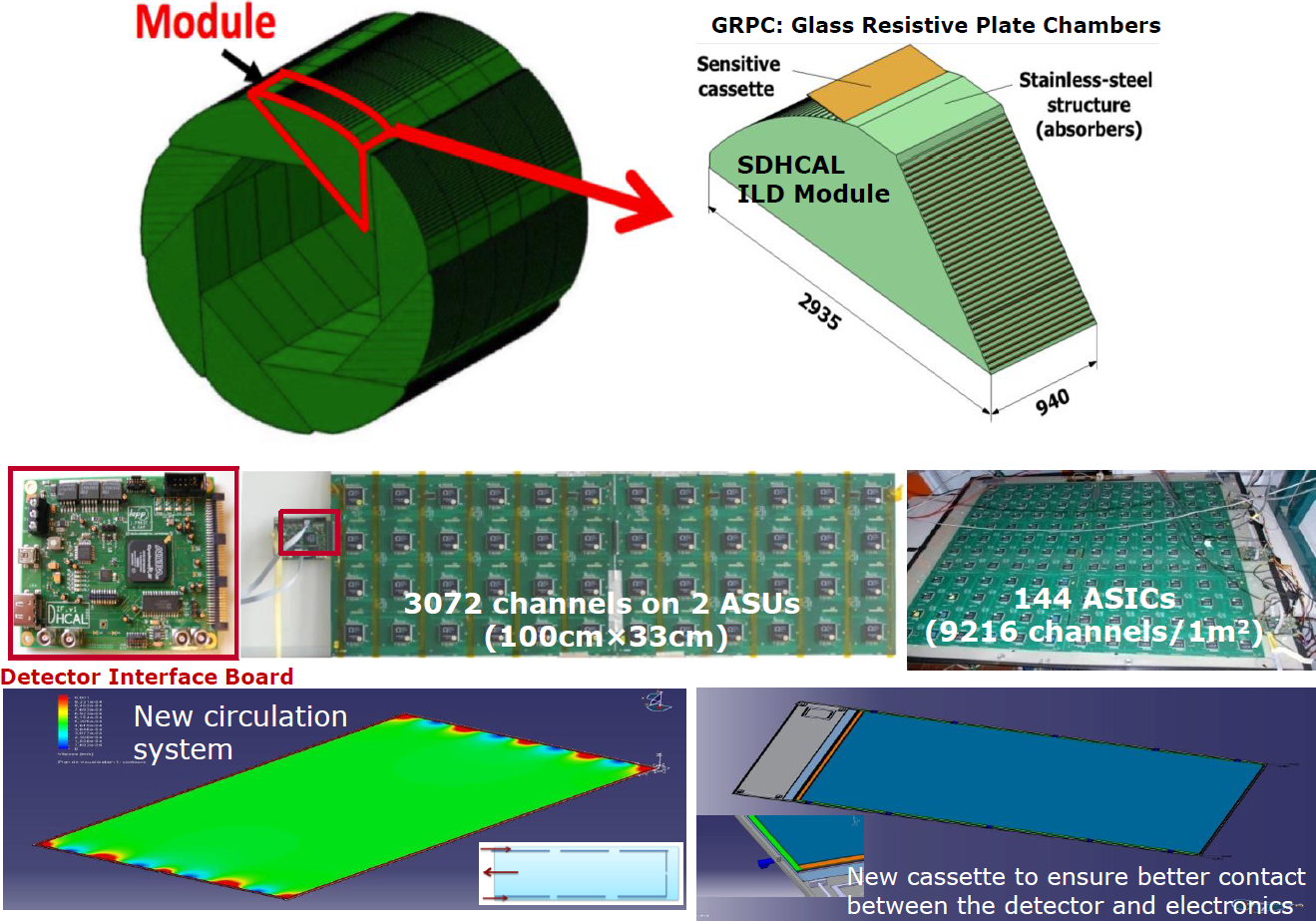}
\caption{\label{fig:SDHCAL_TechProt} The schematics of the SDHCAL layout for the ILC (top) and highlights of the development of the SDHCAL technological prototype (middle and bottom)}
\end{figure}

The layout of the final barrel SDHCAL detector at the ILC is shown in Figure~\ref{fig:SDHCAL_TechProt}, where a barrel module is the aim for the development of the SDHCAL technological prototype. Progress has been made in the new electronics, including the new detector interface board design with smaller dimensions and new readout chips with more features. The development of a large RPC ($3\times1\mathrm{m^2}$) integrated with readout electronics is ongoing. There are also new detector conceptions, e.g. a new gas circulation system to improve the uniformity of gas distribution and a new cassette to ensure better contact between the detector units and electronics. 

\begin{figure}[htbp]
\centering 
\includegraphics[width=.9\textwidth]{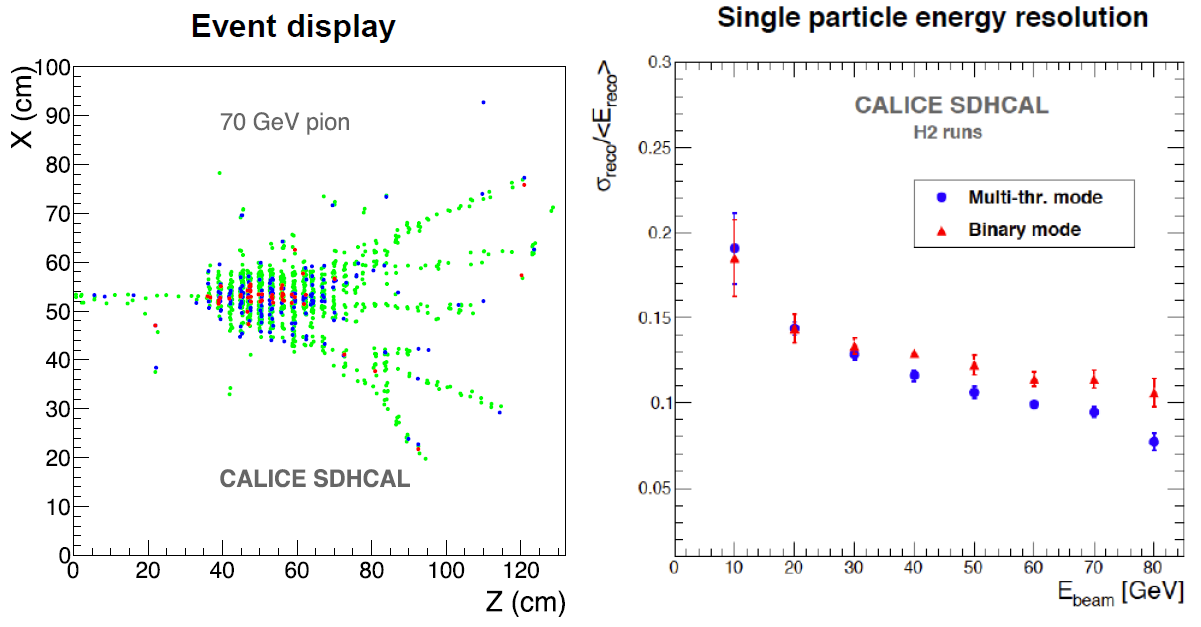}
\caption{\label{fig:SDHCAL_Pion} The event display of 70 GeV pions in the SDHCAL (left) and single-particle energy resolutions of the SDHCAL at different energy points in two readout modes (right)~\cite{SDHCAL_Pion}.}
\end{figure}

Various important tests have been made for the SDHCAL $1\mathrm{m^3}$ prototype using different beams at CERN SPS, including the auto-trigger mode and power-pulse mode. Promising results have been obtained in the hadron shower studies, with highlights shown in Figure~\ref{fig:SDHCAL_Pion}, where the multi-threshold readout mode shows better energy resolutions when the particle energy is above 30 GeV~\cite{SDHCAL_Pion}. 

\section{Other applications of CALICE technologies}
\label{sec:otherapp}
Several detector technologies for the LHC Phase-II upgrade are inspired by the successful CALICE highly granular calorimeters to cope with high pile-up scenarios and harsh radiation environment.

%Desired good spatial resolution would be achievable by the high granularity, good timing resolution is required to separate vertices from different collisions to mitigate pile-up.

The Highly Granular Calorimeter (HGCAL) upgrade of the CMS endcap calorimeter system strongly involves the CALICE-developed technologies~\cite{CMS_Upgrade}. Sectors expected with high-level radiation dose will be equipped with radiation-tolerant silicon sensors and readout chips, similar to the CALICE silicon-tungsten ECAL technology. Other sectors estimated with much less radiation is envisaged to adopt the design of scintillator tiles read out by SMD-SiPMs, which originates from the CALICE AHCAL technology. 

For the ATLAS Phase-II upgrade, a High Granularity Timing Detector (HGTD), inspired by the CALICE silicon-tungsten ECAL technology, is proposed to instrument the gap region between two liquid Argon cryostats~\cite{ATLAS_Upgrade}.

\section{Summary}
\label{sec:sum}
The CALICE collaboration is developing high-granularity calorimeters based on the particle flow approach. Detector concepts have been validated with physics prototypes. CALICE data with different active and passive media have provided possibilities to study hadronic showers in unprecedented granularity and will contribute substantially to further development of hadronic models in Geant4. Technological prototypes are being developed to prove the scalability to construct a full detector. Promising progress has been made including fully integrated electronics, scalable DAQ systems and feasibility of mass production and mass assembly. Further developments are being carried out to address remaining technological challenges. CALICE technologies also find applications in future High-Luminosity LHC experiments.

%\begin{figure}[htbp]
%\centering % \begin{center}/\end{center} takes some additional vertical space
%\includegraphics[width=.4\textwidth,trim=30 110 0 0,clip]{example-image-a}
%\qquad
%\includegraphics[width=.4\textwidth,origin=c,angle=180]{example-image-b}
%% "\includegraphics" from the "graphicx" permits to crop (trim+clip)
%% and rotate (angle) and image (and much more)
%\caption{\label{fig:i} Always give a caption.}
%\end{figure}

%\appendix
%\section{Some title}
%Please always give a title also for appendices.

\acknowledgments

%This project has received funding from the European Union's Horizon 2020 Research and Innovation programme under Grant Agreement No. 654168. 

The author would like to express his gratitude to Dr. Frank Simon, Dr. Marina Chadeeva and other CALICE colleagues; he is also indebted to the INSTR17 Local Organising and Program Committee.

%\paragraph{Note added.} Details on the Si-W ECAL progress should be left to Vladik.

% We suggest to always provide author, title and journal data:
% in short all the informations that clearly identify a document.

\end{document}